# Possible use of a Cooper-pair box for low-dose electron microscopy

Hiroshi Okamoto[1]

*Department of Electronics and Information Systems, Akita Prefectural University*

*Yurihonjo 015-0055, Japan*

A transmission electron microscope that takes advantage of superconducting quantum circuitry is proposed. The microscope is designed to improve image contrast of radiation-sensitive weak phase objects, in particular biological specimens. The objective in this setting is to measure the phase shift of the probe electron wave to a precision $\Delta\theta$ within the number of electrons $N$ that does not destroy the specimen. In conventional electron microscopy $\Delta\theta$ scales as $\sim 1/N^{1/2}$, which falls short of the Heisenberg limit $\sim 1/N$. To approach the latter by using quantum entanglement, we propose a design that involves a Cooper pair box placed on the surface of an electrostatic electron mirror in the microscope. Significant improvement could be attained if inelastic scattering processes are sufficiently delocalized.

[1]Email address: okamoto@akita-pu.ac.jp

# I. INTRODUCTION

Radiation damage governs the image resolution of unstained biological molecules in cryoelectron microscopy [1-3]. The reason is that only a small dose (~ $10^3$ electrons/nm$^2$) of imaging electrons, associated with poor signal-to-noise ratio, is enough to damage the specimen. As a result, the image resolution of a single biological specimen is typically a few nanometers. This falls short of the resolution of 0.8 nm needed to see the secondary structure of a protein; or 0.3 nm needed for fitting the known atomic structures of amino acids to the image. To bypass the problem of radiation damage, several methods have been implemented. First, averaging helps if many copies of the specimen are available. Two-dimensional crystallography [4] and single particle analysis [5] have both been successful in obtaining essentially atomic-resolution structures of proteins in a growing number of cases [6,7]. However, these methods are limited to 'rigid' molecules to ensure uniformity of the molecular structure to be averaged, while many biologically important molecules do not fall into this category. Second, there are ongoing developments in the field of in-focus phase contrast electron microscopy [8-10]. Although impressive contrast enhancement has been demonstrated, it is fair to say that resolution improvement has been modest so far.

There are several theoretical ideas that could prove useful in the future. Quantum state estimation theory [11] tells us that measurement can be done efficiently, i. e. with the minimum dose of electrons in the present case, if performed 'intelligently'. For example, the present author among others proposed to manipulate the scattered electron waves [12-14] to improve information collection efficiency in the context of biological electron microscopy. Another approach, which has been extensively explored in the context of optics [15], takes advantage of quantum entanglement to attain the Heisenberg limit in the setting of parameter estimation.

Related concepts include interaction-free measurement [16], its improvement [17], its application to biological electron microscopy [18]; as well as the repeated use of single photons [19].

The main objective of the present theoretical work is to evaluate the possibility to exploit quantum entanglement in biological electron microscopy. It is natural to consider quantum objects being developed in the field of quantum information processing, as an entity to be entangled with the electron. Though there could be several options, here we consider the Cooper pair box (CPB), which is a superconducting single electron device [20]. A remarkable property of the CPB is that it can be in a superposed state of two macroscopically distinct charge quantum states, as has been demonstrated in 1999 [21]. The proposed microscope uses only a single CPB along with the imaging electron, making it a novel quantum information processing task involving essentially only two quantum objects at any given moment. The CPB interacts with the imaging electron on the surface of a device called electron mirror in our scheme. The electron mirror is an electron optical device that reflects electrons back by an electrostatic field. It has long history of development, beginning in the mirror electron microscopes [22] and the Castaing-Henry type energy filters [23]; and more recently in the low energy electron microscope (LEEM) in the mirror mode [24] and the mirror-based aberration correctors [25,26]. Additionally, several proposed schemes for biological electron microscopy involve an electron mirror with a micro-fabricated structure [13,14]. The present scheme involves a very-low temperature electrostatic electron mirror [27] in the electron microscope, in which the electrons interact electrostatically with the CPB placed near the turning point of the electron beam on the electron mirror surface. Since the electrostatic field around the CPB can be in a quantum superposition, the electron-CPB interaction should exhibit unusual properties. We will describe how to use such properties to improve electron microscopy of radiation-sensitive biological molecules. We will also need a well-controlled pulsed single electron source [28], since the timing control of electron emission at the electron source is essential for our scheme to work.

The rest of the paper is organized as follows. In Section II, we review some basics; and then discuss phase shift measurement methods that attain either the standard quantum limit or the Heisenberg limit. In Sec. III, we illustrate the main idea behind our proposed scheme. This is followed by Section IV, in which we present in detail the structure of the proposed electron microscope, the basic quantum operations involving both the electron and the CPB, as well as practical considerations. In Section V, we present possible measurement protocols along with a simulation study. Section VI concludes the paper. Throughout the paper, we assume the paraxial ray approximation to be valid. The term *diffraction plane* will be used for any plane conjugate to the back focal plane of the objective lens. Symbols $e$, $m$, $k_B$ and $h$ represent the proton charge, electron mass, Boltzmann's constant and Plank's constant, respectively. In light of the recent development of aberration correction techniques [29], and also due to the relatively low resolution to be dealt with, we take the liberty of not explicitly considering geometrical aberrations in electron optics.

## II. PHASE MEASUREMENT AND THE HEISENBERG LIMIT

We first review the limit of conventional cryoelectron microscopy. For this purpose, we consider the best case, which is in-focus phase contrast imaging [8-10] with an ideally working phase plate and an electron detector limited only by shot noise. The unstained biological specimens behave as weak phase objects. What we want to measure is the phase shift $\theta$ of the electron wave at the pixel of interest, relative to the phase shift associated with the whole transmitted electron wave, which experiences the average phase shift over the whole illuminated portion of the specimen. Since we have in-focus phase contrast, the small phase shift $\theta$ manifests itself as intensity modulation $|1 + \theta|^2 \sim 1 + 2\theta$ on the corresponding pixel of the electron detector. Let us assume that we are interested in determining the phase shift $\theta$ to a precision $\Delta\theta$. Let $N$ be the expected number of electrons that a single pixel receives. According to Poisson statistics, the actual electron count has the standard deviation $N^{1/2}$. Thus, we obtain the relation $\Delta\theta \sim N^{1/2}/2N = 1/2N^{1/2}$ between the precision of phase shift measurement $\Delta\theta$ and the necessary dose $N$. This inverse-square-root dependency is known as the *standard quantum limit* or *shot-noise limit*. All methods including e. g. holography [30] that use electrons independently are governed by the standard quantum limit.

To discuss attainable image resolution based on the measurement precision $\Delta\theta$ of the phase shift, it is unavoidable to make rather shaky assumptions about the *specimen* [14]. First, observe that the structure of a natural specimen is not random: One should expect similar thicknesses at two nearby points on a natural, biological object. As the separation $l_1$ between the two points gets shorter, the thickness difference we need to detect is expected to be smaller too, demanding a larger number of electrons. To take this into account, we assume that the expected difference $\Delta\theta$ of two phase shift values for two points separated by a sufficiently small $l_1$, which we assume to be positive without loss of generality, is proportional to $l_1$;

$$\Delta\theta \sim \alpha l_1. \qquad (1)$$

The difference of the phase shift per unit travel length, between electron waves passing through typical biological matter and the surrounding vitreous ice, is of the order 10 mrad/nm. Hence it is natural to take a value $\alpha \sim 10$ mrad/nm as the proportionality constant [14]. Second, experiences seem to be consistent with the idea that electron irradiation induces destruction of specimen features finer than the length scale

$$l_2 \sim \gamma n, \qquad (2)$$

where $\gamma \sim 10^{-3}$ nm$^3$ is a constant and $n$ is the electron dose per unit area [2,14]. Now, consider a square pixel with the side length $l$ on the specimen plane, through which a dose of $N = nl^2$ electrons cross. A condition $l_1 = l_2 = l$ then means that the pixel size corresponds to the smallest possible resolution. Together with the relations $\Delta\theta \sim 1/N^{1/2}$ and $N = nl^2$, we obtain a resolution value

$$l \sim (\gamma/\alpha^2)^{1/5} \sim 1.6 \text{ nm}. \tag{3}$$

This result, considering the crudeness of the assumptions we made, is not too far from the experimental resolution 4-5 nm of cryoelectron microscopy of single objects.

The attainable resolution in the above case is determined by the standard quantum limit. The objective of the present work is to move towards the *Heisenberg limit* $\Delta\theta \sim 1/N$. Let us first see the consequences of such a phase measurement partially governed by the Heisenberg limit, regardless of how it can be done. Suppose that $k$ electrons can be used, as a whole, to perform phase measurement at the Heisenberg limit. More explicitly, we consider a situation in which the group of $k$ electrons can be seen as a single hypothetical particle with the following properties: 1) It damages the specimen as much as $k$ electrons combined. 2) At the same time, it receives a $k$-times larger phase shift $\theta' = k\theta$ upon crossing the specimen. The number of the hypothetical particles that we can use is $N' = N/k$. On the other hand, since $\Delta\theta' \sim 1/N'^{1/2}$ we obtain a relation $\Delta\theta \sim 1/(Nk)^{1/2}$. (In particular, we have $\Delta\theta \sim 1/N$ if $k = N$.) This is a key point and it is worth mentioning another perspective: Shot noise $1/N'^{1/2}$ actually *increases* by a factor $k^{1/2}$ because of the reduced number of 'hypothetical particles' compared to the number of electrons. At the same time, however, we have an enhanced signal, or contrast, by a factor $k$ because of the relation $\theta' = k\theta$, leading to an improved signal-to-noise ratio proportional to $k^{1/2}$. Recall also the so-called Rose inequality $C > 5/N^{1/2}$ [31], which states a condition to detect an image feature with a contrast $C$ by using $N$ probe particles crossing the feature. (The numerical factor 5 here reflects a typical desired confidence level when recognizing a feature in a shot-noise dominated image.) While $C$ on the left hand side of the equation is proportional to $k$, the term $5/N^{1/2}$ on the right hand side is proportional to $k^{1/2}$; thus one can always satisfy the Rose inequality if a large enough $k$ is employable.

The relation $\Delta\theta \sim 1/(Nk)^{1/2}$ leads to a modification to the expression Eq. (3) of the attainable resolution, resulting in

$$l \sim (\gamma/k\alpha^2)^{1/5} \sim 1.6 \text{ nm}/k^{1/5}. \qquad (4)$$

Hence, a resolution improvement by a factor 2 would require $k = 32$. In the extreme case of $k = N$, the attainable resolution is $l \sim (\gamma/\alpha)^{1/4} \sim 0.6$ nm, which requires $k \sim 170$. We will see in later sections that a measurement with a large $k$ is increasingly difficult to realize, besides the (possibly tractable) problem of having $k\Delta\theta > 2\pi$. Hence 0.6 nm may be regarded as a lower-bound, which presumably is not tight, for the attainable resolution in terms of length. Despite the uncertainties about the parameter values $\alpha$ and $\gamma$, and the validity of equations (1) and (2) for that matter, these arguments suggest that the proposed method alone is unlikely to deliver atomic resolution.

One way to grasp the phase measurement described above may be to consider a single electron passing the specimen $k$ times [14] with the exit wave refocused on the specimen plane each time; thus the electron receives a phase shift $k\theta$ while damaging the specimen $k$ times, before going to the subsequent phase plate, followed by projector lenses and the screen. Hence, this process may be regarded as the 'hypothetical particle' described above crossing the specimen once. However, the reader should be warned that such a scheme would be impractical because of the complicated electron optics, the wide spread of the electron beam due to repeated scattering, and the exponential loss (in terms of $k$) of electrons due to inelastic scattering, which is fundamentally unavoidable in cryoelectron microscopy. Indeed, inelastic mean free path of 120 keV electrons in vitreous ice is known to be ~ 160 nm [32] while the size of various interesting biological structures can be more than 50 nm; resulting in significant probability of inelastic scattering for each pass. Hence we need a measurement scheme that is resilient against lossy inelastic processes. In the next section, we describe one candidate scheme for this purpose.

**III. THE MAIN IDEA**

In this section we discuss the main idea, deferring considerations on electron optics and feasibility to later sections. Figure 1 shows the configuration of the proposed microscope at the conceptual level. Electron emission at the electron source ES is followed by reflection (ideally on a diffraction plane) by the electron mirror EM with a Cooper-pair box CPB. Subsequently, the electron passes the specimen SP placed on an image plane and then detected by an area detector AD placed on a diffraction plane. This microscope does not produce the entire image at once. Rather, it measures the difference $\varDelta\theta$ of two phase shifts $\theta_0$ and $\theta_1$, associated with two specimen regions S0 and S1. These regions are small enough and have no significant 'internal structures' (See discussions below). The final phase shift map of the specimen is obtained by a multitude of such phase shift difference measurements. An example will be presented in later sections.

The electron mirror is a device that reflects electrons back by an electrostatic field. In the proposed microscope, the turning points of the electrons are set in the vicinity of the physical mirror device, which is placed in a very low temperature environment to make the superconducting CPB work. The CPB in this setting is either neutral or charged by a single excess Cooper pair. This results in the modulation of the electron mirror surface. When the CPB is not charged, the reflected electron wave is focused on the region S0; while if charged by a single Cooper pair, the reflected wave goes to the region S1. Though the CPB is a micro-fabricated device with limited precision, the mirror surface remains highly smooth when placed slightly away from the physical CPB. In Fig. 1, the two possible mirror surfaces are shown as two dotted lines. The scattered waves from the specimen then reach the area detector, which has a single electron detection capability [33,34].

The remarkable fact is that the CPB is a quantum object, which can be in a *superposed quantum state* of the two basis states; i. e. the neutral state and the charged state [21]. Hence the electron can be focused on the regions S0 and S1 *simultaneously*, which however is not electron wave delocalization in the usual sense. Thanks to the research activities in the field of quantum information processing, the quantum state of the CPB can be initialized, read out, and manipulated at will by using e. g. microwave pulses [35]. In our scheme, microwave irradiation to the CPB can be avoided during the interaction period with the reflecting electrons, so as not to disturb the electron trajectory.

Here we sketch how the microscope works. For wider accessibility, we use wavefunctions to represent quantum states in this section. Since our scheme involves two quantum objects, namely the CPB and the imaging electron, we must use the joint wavefunction $\Psi(\mathbf{r}, i)$ to represent the state of the whole system, where $\mathbf{r}$ is the position of the electron; and $i$ is an integer taking only two values 0 and 1, which represents the CPB state (0: neutral, 1: charged). First, we set the CPB 'wavefunction' $\psi_{\mathrm{CPB}}(i)$ to be $\psi_{\mathrm{CPB},1}(i) = (1/2)^{1/2}\{\delta_{i,0} + e^{i\sigma}\delta_{i,1}\}$, where $\delta_{i,j}$ is the Kronecker's delta that is a discrete analog of Dirac's delta function $\delta(\mathbf{r})$; and $e^{i\sigma}$ is a phase factor for some $\sigma$. Hence the CPB is in an equally superposed state of neutral and charged states. When the electron source emits an electron in the state $\psi_{e,1}(\mathbf{r})$, the whole system is in a product state

$$\Psi(\mathbf{r}, i) = \psi_{\mathrm{CPB},1}(i)\psi_{e,1}(\mathbf{r}). \tag{5}$$

Upon reflection, however, the CPB state and the electron state get *entangled* (i. e., can no longer be factored as in equation (5)) as

$$\Psi(\mathbf{r}, i) = (1/2)^{1/2}\{\delta_{i,0}\,\psi_{e,\mathrm{S0}}(\mathbf{r}) + e^{i\sigma}\delta_{i,1}\,\psi_{e,\mathrm{S1}}(\mathbf{r})\}, \tag{6}$$

where $\psi_{e, S0}(r)$ is the electron wave that passes the S0 region of the specimen and likewise for $\psi_{e, S1}(r)$ passing the S1 region. Now, suppose that the S1 region of the specimen induces a larger amount of phase shift than the S0 region, by an angle $\Delta\theta$. Hence, upon transmission through the specimen, the state becomes, up to the overall phase factor

$$\Psi(r, i) = (1/2)^{1/2} \{\delta_{i,0} \psi_{e, S0}(r) + e^{i(\sigma + \Delta\theta)} \delta_{i,1} \psi_{e, S1}(r)\}. \tag{7}$$

Finally, we detect the scattered electron by the area detector. Let $\psi_{e, ADj}(r)$ be the wavefunction representing the electron wave converging to the $j$-th pixels on the detector. The wave function $\psi_{e, S0}(r)$ can then be expanded as a superposition of partial waves $\psi_{e, ADj}(r)$ as

$$\psi_{e, S0}(r) = \Sigma_j a_j \psi_{e, ADj}(r), \tag{8}$$

where coefficients $a_j$ depends solely on the structure of electron optics. Likewise, we define another set of coefficients $b_j$ by a similar relation $\psi_{e, S1}(r) = \Sigma_j b_j \psi_{e, ADj}(r)$. Rewriting equation (7), we obtain

$$\Psi(r, i) = (1/2)^{1/2} \Sigma_j \{a_j \delta_{i,0} + b_j e^{i(\sigma + \Delta\theta)} \delta_{i,1}\} \psi_{e, ADj}(r). \tag{9}$$

Now, suppose that we detected the electron at the $j$-th pixel. This would leave the CPB state in

$$\psi_{CPB, 1}(i) = (1/2F)^{1/2} \{a_j \delta_{i,0} + b_j e^{i(\sigma + \Delta\theta)} \delta_{i,1}\}, \tag{10}$$

where $F$ is a normalization factor. Since the area detector AD is placed on a diffraction plane and the small regions S0 and S1 are rather structureless, the intensity (i. e. squared amplitude) of both the wave functions $\psi_{e, S0}(r)$ and $\psi_{e, S1}(r)$ spread on the entire detector plane rather uniformly. This means that the coefficients $a_j$ and $b_j$ for the same $j$ has similar amplitudes but differing phase values. This allows us to write $b_j \sim a_j \exp(i\beta_j)$ where $\beta_j$ is a phase angle that depends on $j$. More intuitively, $\beta_j$ is associated with the optical path length difference from the two regions S0 and S1 to the $j$-th pixel of the area detector. Hence equation (10) may be rewritten as, up to the overall phase factor,

$$\psi_{CPB, 1}(i) = (1/2)^{1/2} \{\delta_{i,0} + e^{i(\sigma + \beta_j + \Delta\theta)} \delta_{i,1}\}. \tag{11}$$

We know $\beta_j$ because we know the pixel in which the electron was detected. This allows us to manipulate the CPB quantum state by a microwave pulse to nullify the phase factor $\exp(i\beta_j)$ to obtain

$$\psi_{\text{CPB}, 1}(i) = (1/2)^{1/2} \{\delta_{i,0} + e^{i(\sigma + \Delta\theta)} \delta_{i,1}\}. \tag{12}$$

As will be seen in Sec. V, this manipulation may be postponed to the final stage of the measurement, meaning that the operation does not have to be done every time an electron is detected.

The overall effect is an addition of the phase shift $\Delta\theta$ to the phase angle $\sigma$ contained in the initial wavefunction of the CPB. By repeating the above process, the phase shift $\Delta\theta$ *linearly accumulates* in the CPB quantum state, as in the 'hypothetical particle' situation discussed in the final part of Sec. II. Thus, the above process, repeated $k$ times, allows us to approach the Heisenberg limit. After the repetition, the CPB quantum state is manipulated and measured by methods developed in superconducting quantum circuit technology. The measurement yields 1 bit of information about the accumulated phase shift $k\Delta\theta$. Note, however, that we did not discuss the resiliency of the scheme against inelastic scattering processes. For example, the entangled state (6) will be destroyed if the electron state is effectively projected to a localized state in the region S0 or S1 upon inelastic scattering. We will consider such lossy processes in Sec. V.

**IV. THE PROPOSED ELECTRON MICROSCOPE**

In this section, we describe the 'actual' structure of the proposed microscope. As above, the incident electron waves are directed either on the specimen region S0 or S1, and the CPB can take only two states. Hence essentially we are dealing with two state quantum systems, or *qubits* in the terminology of quantum information science, where the CPB is being extensively studied.

Following the convention in the field, hereafter we represent the electron waves directed to the regions S0 and S1 by vectors |0> and |1>, respectively.

The main idea discussed in the previous section should be applicable to various actual microscope designs. In particular, the regions S0 and S1 need not be two adjacent regions with a similar size. Hereafter we consider a particular configuration in which S0 is a relatively large circular region (with a diameter ~ 3 nm for instance), while S1 is a small circular region (with the size of desired image resolution, i. e. diameter ~ 0.5 nm) placed at near the center of S0. Our intention is to measure the phase shift at the 'pixel' S1, relative to the average phase shift over the larger area S0 that surrounds the pixel, resembling in-focus phase contrast microscopy. The whole map of the specimen can then be acquired by raster-scanning both the regions S0 and S1, while keeping their relative position constant. This will give us a kind of differential phase image, which is a map of protrusions and depressions in the phase shift map. As discussed later, the reason why we cannot make the region S0 larger (as in conventional phase contrast microscopy) is that the size of the region S0 should be smaller than inelastic scattering delocalization length, which is considered to be a few nm [36].

## A. The structure of the microscope

Figure 2 shows the proposed TEM. Probe electrons are emitted at desired times from a pulsed electron source (PES) that is similar to the one described in Ref. [28]. Ideally, one electron interacts with the specimen at a time. However, our scheme is resilient against failures to generate electron pulses or to generate pulses containing multiple electrons, but the pulses should not be generated at unintended times. A monochrometer (MC) follows to produce an electron beam with an energy spread of 0.1-1 meV, a requirement that will be discussed later. Although a monohromator that has been implemented so far in a TEM has energy width of ~ 60 meV [37], we note that energy spread of 10-30 meV in TEM is being discussed [38] and a monochrometer with 1.2 meV energy half width (full width at half maximum) has been reported in the context of low energy electron energy loss spectroscopy [39]. (In our scheme, all components on the left of the condenser lens (CL) in Fig. 2, excluding the CL itself, are held at a similar electrostatic potential. It is only in this potential region we need to maintain the stability of electrostatic potential differences down to 0.1-1 meV. Since relative potential differences among these components are small, low energy techniques are indeed relevant. For the sake of concreteness, hereafter we assume that the electron kinetic energy is ~ 10 eV in this part of the instrument so that we have $\Delta E/E < 10^{-4}$.) Subsequently, the electron mirror (EM) combined with an electron beam separator (EBS) follows [23-26]. The electron beam separator uses a magnetic field to direct the incident electron beam to the mirror and guide the reflected electron beam from the mirror to the output port (i. e. the right-hand side of EBS shown in Fig. 2). The CPB on the surface of the electron mirror modulates the mirror surface shape depending on the charge state of the CPB. Since the CPB should be held at a very low temperature of the order of ~ 10 mK, the electron mirror device will have to be held most likely by a dilution refrigerator and the electron beam path to it must be so configured that room temperature, or 78K radiation, does not impinge on the mirror surface with too large an associated solid angle. The use of other quantum devices (e. g. quantum dots) working at higher temperatures may potentially relax some of these conditions. Then, the electrons are accelerated to the imaging energy of ~ 100 keV. In the present scheme, presumably all the components before the CL (i. e. PES, MC, EBS, EM and CPB) should be held together at a high negative electrostatic potential on the order of -100 keV, relative to the rest of the microscope. Such a configuration may seem far from conventional, but there is no fundamental problem because a group of electronics components may be floated together [40]. In particular, the 0.1-1 mV level stability, which is required for producing the entangled quantum state, is *not* required for the ~ 100 kV acceleration voltage source. This situation is unlike the TEM-EELS, where the high voltage source needs to have the stability comparable to the desired energy resolution. Subsequently, the condenser lens (CL), specimen, and objective lens (OL) follow in this order. Finally, a system of projector lenses (PLS) and the final position-sensitive electron detector (ED) is in place to measure the scattered electrons. The electron detector ED should be able to detect single electrons with high efficiency [33]. Note that "approximately 100%" electron detection efficiency has already been demonstrated in 1989 [34].

**B. A brief review of CPB physics**

We provide a brief review of the CPB and some definitions for later sections. Figure 3 (a) shows a representative circuit containing a CPB with associated circuit parameters. CPB circuits are usually fabricated by electron-beam lithography and the lateral size of a CPB is typically 0.1-1 μm. Aluminum, with the superconducting gap energy $\Delta$ = 180 μeV (while somewhat larger values are usually found in microfabricated devices), is usually the material of choice because of the good quality of oxide films employed as the insulating barriers of tunnel junctions. The CPB in Fig. 3 (a) is connected to the ground electrode through a Josephson junction (with the junction capacitance $C_J$ and the Josephson energy $E_J$) and therefore Cooper pairs can go in and out of the CPB. Another electrode, the bias electrode at the potential $V_g$, is coupled to the CPB via the capacitance $C_g$. Let the excess number of Cooper pairs on the CPB be $n_C$. The electrostatic energy of the system is

$$(-2n_C e + C_g V_g)^2/(2C_\Sigma), \qquad (13)$$

where $C_\Sigma = C_J + C_g$ [20]. The CPB is made sufficiently small (with the typical lateral size less than ~ 1 μm) and only a single excess Cooper pair in the CPB can charge the CPB significantly with energy $4E_C$ when $V_g$ = 0. The charging energy $4E_C$ (where $E_C = e^2/(2C_\Sigma)$) should be smaller than the thermal energy $k_BT$, for the CPB to work as a 'single electron device'. In what follows, we assume the value $E_C \sim 100$ μeV. We operate the CPB in the region of well-defined number of Cooper pairs, i. e., $E_J \ll E_C$. Then, the operation temperature $T$ must satisfy $k_BT \ll E_J$. We assume parameters $E_J \sim 10$ μeV and $k_BT \sim 1$ μeV in the following considerations, because the former corresponds to a realizable Josephson critical current of ~ 5 nA while the latter temperature $T \sim 10$ mK can be reached by the $^3$He/$^4$He dilution refrigerator. Under these conditions, the excess charge on the CPB is zero. Let this charge state of the CPB be $|0>_b$ and the state with exactly one excess Cooper pair on the CPB be $|1>_b$ (The subscript $b$ stands for 'box' to indicate that the state is not about the vacuum electron but about the CPB). Suppose, from now on, that the bias voltage is set to be $V_g = e/C_g$ (henceforth referred to as the *charge degeneracy point*). Then, equation (13) tells us that the two states $|0>_b$ and $|1>_b$ are equally favorable energetically as far as electrostatics is concerned. Intuitively, a Cooper pair 'wants' to enter the CPB through the Josephson junction because of the positive bias voltage $V_g$, but that would entail negative charging of the CPB, causing frustration. The electrostatic potentials, with respect to the ground electrode, associated with these two states are different: They have the same magnitude $e/C_\Sigma \sim 200$ μV but have opposite signs, and hence the difference is $\Delta\varphi \sim 400$ μV. The weak Josephson energy $E_J \ll E_C$ comes into play at this point. In the region of well-defined charge that we work in, the Josephson energy part of the Hamiltonian takes the form [35]

$$- (E_J/2)(|0>_b<1|_b + |1>_b<0|_b). \tag{14}$$

At the charge degeneracy point, this is the dominating part of the Hamiltonian and hence the energy eigenstates are superposed states $|s>_b = (|0>_b + |1>_b)/2^{1/2}$ and $|a>_b = (|0>_b - |1>_b)/2^{1/2}$, which are separated by the energy gap $E_J$. Several experimental groups have controlled such a CPB coherently and read out the associated quantum state [21]. For example, realizable quantum operations on the CPB includes $|0>_b \rightarrow |0>_b$ and $|1>_b \rightarrow e^{i\kappa}|1>_b$ for an arbitrary $\kappa$. This operation will be referred to as the *phase shift operation* with an angle $\kappa$. See Ref. [35] for a detailed account of some representative methods for these operations.

**C. The CPB placed on the electron mirror**

Here we mention the overall scheme of the electron mirror equipped with the CPB. We defer practical considerations to the subsection D. The mirror consists of a microfabricated *mirror device* with a CPB and the *aperture electrode* that has a small aperture as shown in Fig. 3 (b). The aperture electrode need not be integrated to the mirror electrode with microfabrication techniques because these are likely to be separated by millimeters. The potential difference between the *mirror electrode* on the mirror device and the aperture electrode is such that the electrons are repelled by the electric field $E_M$. The electron energy is such that the turning point of an electron trajectory is set close (~ 1 μm) to the mirror device surface. On the other hand, all the electrodes in the mirror device have similar potentials typically within millivolts. The mirror surface, as opposed to the physical surface of the mirror device, is shown as two dotted lines M0 and M1: The lower surface M0 corresponds to the CPB state $|0>_b$ that lacks an excess Cooper pair. Likewise, the upper surface M1 corresponds to the negatively charged CPB state $|1>_b$. The base electrode to which the CPB is connected through a Josephson junction is independently biased, but at a potential close to the bias electrode. The exact potential difference between the bias electrode and the base electrode is such that the CPB is set at the charge degeneracy point. Additionally, the potential of the mirror electrode is adjusted relative to the base electrode so that the surface M0 is approximately flat.

Figure 3 (c) shows a schematic drawing of the electron optics. The electron gun and the associated lenses produce an effective electron point source at the point A near the electron mirror surface, at the location of the CPB. The incident electron wave from A (represented by $|i>$) is then reflected to produce either the wave state |0> by the mirror surface M0, or the state |1> by the surface M1. Acceleration of the reflected electrons results in curved trajectories. Since the surface M0 is planer and M1 is convex, the electron waves emanating from the mirror can be represented by two virtual point sources VS0 and VS1. (Here we do not study whether these virtual point sources are good enough point sources. We only point out that deviations from such ideality could be corrected either by hardware or software means.) The 'lens system' in Fig. 3 (c) schematically represents the combination of *all* lenses between the mirror device and the specimen that include the aperture lens associated with the aperture electrode. The electron kinetic energy may well change within the lens system. The overall effect is that the electron wave in the state |0> illuminate the large area S0, while the wave in the state |1> focuses on a small region S1 on the specimen, as desired. Thus, the action of the electron mirror is therefore summarized as

$$|0>_b|i> \rightarrow |0>_b|0>, \quad |1>_b|i> \rightarrow |1>_b|1>. \quad (15)$$

After scattering by the specimen, the electron is detected at a plane conjugate to one of the 'Detection' planes shown in Fig. 3 (c), after suitable magnification by the post-specimen lens system beginning with the objective lens. The places of the 'Detection' planes are where the rims of the two rays corresponding to the states |0> and |1> meet, because *we do not want to know* whether the electron was in the state |0> or |1> before scattering. Knowing the state would amount to knowing the charge state of the CPB, thus destroying the superposed CPB state of equation (12). Put another way, the amplitudes of the coefficients $a_j$ and $b_j$ for a given pixel *j*, mentioned in Sec. III, should be kept as close as possible, in order not to get much information about whether the state was |0> or |1>.

**D. Practical considerations**

Here we consider three experimental issues in a semi-quantitative manner. For simplicity, we fix the imaging electron energy to be 100 keV. We use the term *propagation angle* to mean the angle between the optical axis $z$ and the momentum vector $\boldsymbol{p}$ of the electron.

First, we examine the electron optics shown in Fig. 3 (c). Let us study whether the tiny potential change, on the order of $\Delta\varphi \sim 400$ μV, of the CPB is enough to cause the intended change of the incident electron wave to the specimen. Specifically, the CPB state switches the focal point between the points P and Q shown in Fig. 3 (c). While point Q is on the specimen plane, the point P should be sufficiently far from the specimen plane so that the electron wave incident on the large region S0 is approximately a plane wave. In other words, when the separation $d$ between the points P, Q is too small, we get an undesirable shadow image in the far field that is essentially an image of the specimen with a defocus $\sim d$. As mentioned in Sec. III regarding equation (8), large amplitude variations among the parameters $a_j$ should be avoided in our scheme. We assume that $d \sim 1$ μm defocus is good enough for our purpose because the region S0 is so small (3 nm) that the associated half angle is then $\sim 1.5$ mrad.

We assume that the lens system shown in Fig. 3 (c) has the overall magnification $M \sim 1/200$, so that a region of a size $\sim 100$ nm on the CPB is focused to the region S1 with the size $\sim 0.5$ nm. (Note that the electron motion perpendicular to the optical axis is not affected by the field $\boldsymbol{E}_M$ along the $z$-axis as long as $\boldsymbol{E}_M$ is uniform.) As the region S1 associated with point Q has a spatial extent $\sim 0.5$ nm, the diffraction effect demands us to let the propagation angle to the region S1 be larger than $\sim 10$ mrad. Hence, the corresponding propagation angle $\beta$ on the electron mirror side of the lens is $\beta \sim 10$ mrad/$M \sim 0.05$ mrad. This means that the energy associated with the electron motion perpendicular to the optical axis can be as large as $\sim 100$ keV $\beta^2 \sim 300$ μeV. Equivalently, the electron kinetic energy $E_z = p_z^2/2m$ along the optical axis $z$, where $p_z$ is the momentum along the optical axis, has the associated energy spread ~ 300 μeV even if the electron beam were perfectly monochromatic. We must ensure that all electrons with various propagation angles get close to the CPB to 'feel' the electrostatic field generated by the CPB. Hence two equipotential surfaces near the CPB with a potential difference ~ 300 μV should be within the length scale comparable to the size $l_{CPB}$ ~ 0.1 μm of the CPB. Thus we have the desirable size of the field $E_M = |\boldsymbol{E}_M|$ > 3 mV/μm ~ 3 kV/m near the mirror surface. This argument also makes it clear that a highly monochromatic electron beam with the energy spread $\Delta E$ on the order of 0.1-1 meV is needed. The propagation angle gets larger near the mirror surface because of deceleration, by a factor proportional to square root of the ratio of the initial and final electron energy. In particular, the propagation angle is ~ $\beta$ (100 keV/1 meV)$^{1/2}$ ~ 0.5 rad near the mirror surface. In other words, the CPB need to exert a force to the electron so that the propagation angles of the reflected electrons changes by ~ 0.5 rad depending on the CPB state. Hence the mirror surface M1 should have angle ~ 0.25 rad from the 'flat' M0 surface, which should be realizable by a proper electrostatic configuration.

On the other hand, too large an $E_M$ would make the small CPB potential 'undetectable'. More precisely, in order to get a considerable momentum in parallel with the mirror surface upon reflection, the electron wave should have a significant phase change along $z$-direction. The electron 'wavelength' $D$ at the mirror surface is estimated to be, based on the WKB approximation, $D \sim (h^2/meE_M)^{1/3}$ [14]. For the CPB-generated potential to make a 'bump' on the mirror surface larger than the size $D$, a condition $\Delta\varphi > DE_M$ should be satisfied. Thus, we have $E_M < (me\Delta\varphi^3)^{1/2}/h$ ~ 5 kV/m that, together with the preceding condition, determines the desirable $E_M$ to be around 1-10 kV/m. Note that this value is much smaller than the known practical upper limit ~ $10^7$ V/m due to electrostatic breakdown in a vacuum.

Second, we address the timing-related issue. Note that, in the presence of nonzero Josephson energy, the CPB state goes back and forth between the states $|0>_b$ and $|1>_b$ with the frequency $E_J/h \sim 1$ GHz. Hence the electron arrival times must be controlled to a precision better than 1 ns and the pulsed electron source should be controlled to this precision. Moreover, the electrons must travel in the microscope with a predictable time period $\tau = L/v \sim 500$ ns, where $L$ and $v$ are the characteristic size of the electron microscope ($\sim 1$ m) and the velocity ($\sim 2 \times 10^6$ m/s) of the electrons with $\sim 10$ eV kinetic energy. The 10 eV energy is the characteristic energy scale in the portion of the microscope, which generates the entangled electron state. Roughly, the broadening $\Delta\tau$ of the time period $\tau$ due to the electron beam energy spread is given by $\Delta\tau/\tau \sim \Delta E/E < 10^{-4}$ for energy larger than 10 eV, and hence we obtain $\Delta\tau < 1$ ns, which ensures successful operations of the electron mirrors. We also touch on the lifetime of the CPB quantum states. A CPB qubit lifetime of > 2 µs has been described as "a worst case estimate" consistent with previous experiments [35]. On the other hand, Ref. [28] reports a pulsed electron source in a TEM that employs sub-100-fs optical pulses with a 80 MHz repetition rate. Hence, these experimentally demonstrated parameters suggest that $\sim 160$ electron pulses may be generated within the lifetime of the CPB quantum state. An additional time scale to consider is the image acquisition time. For example, the simulated images described in Sec. V, B are generated by detecting $1.8 \times 10^4$ simulated electrons. If we assume that the electron detection rate is the above value of $\sim 80$ MHz (corresponding to a $\sim 10$ pA current), then it would take $\sim 0.2$ ms to form an image. However, the electron detection rate could be made much lower if necessary. For example, an electron detection rate as low as $\sim 10$ Hz has been reported [34].

Third and finally, we consider the following question: Besides the natural CPB oscillation $|0>_b \leftrightarrow |1>_b$ at the frequency $E_J/h$, does the reflecting electron 'push' the Cooper pair contained in the CPB to trigger transitions such as $|1>_b \rightarrow |0>_b$, thereby rendering the equation (15) inaccurate? The short answer is that, since the CPB state dynamics is such that it cannot change faster than the characteristic time scale $h/E_J \sim 1$ ns, the 'sudden approximation' of quantum mechanics states that the number of Cooper pairs in the CPB remains the same if the electron reflection process is fast enough. It is indeed fast as long as the following two time scales are sufficiently small compared to $h/E_J$: The first time scale $\tau_1 \sim h/\Delta E$ is derived from the electron beam energy spread $\Delta E$. A suitable $\Delta E \sim (0.1\text{-}1)$ meV that fulfills $E_J << \Delta E \sim e\Delta\varphi$ can be chosen because of the condition $E_J << E_C$ in our scheme. The second time scale is $\tau_2 \sim (ml_{CPB}/eE_M)^{1/2} \sim$ 10 ps, during which the electron 'feels' the electrostatic force by the CPB. Note that $\tau_2$ is a time period for the reflecting electron with the acceleration $eE_M/m$ to move the distance $l_{CPB} \sim 0.1$ μm, which is roughly the size of the CPB. If needed, the use of a Josephson junction with a smaller $E_J$ would result in a wider margin to meet the above conditions. This, however, would entail a lower operation temperature.

Here we elaborate on the above argument. Figure 3 (d) depicts another way to look at the electron mirror device. The electron wave normal to the surface of the mirror is shown as the state |0>, while the wave with an oblique angle is shown as |1>. When the state of the CPB is |0>$_b$, upon reflection on the mirror surface M0 the electron state remains the same. In contrast, if the CPB state is |1>$_b$, the electron state flips as |0> ←→ |1>. In quantum information science terminology, this is nothing but the controlled-not (CNOT) gate, in which the CPB state controls the electron state. We define, like in the CPB case, the symmetric electron state |s> = (|0> + |1>)/$2^{1/2}$ and the asymmetric state |a> = (|0> - |1>)/$2^{1/2}$. It is known that in this alternative basis, the CNOT state becomes another CNOT state, in which the electron state {|s>, |a>} does control the flipping of the CPB state in the basis {|s>$_b$, |a>$_b$}. Hence the electron and the CPB exchange energy $E_J$ because the CPB states |s>$_b$ and |a>$_b$ have the energy difference $E_J$. However, such an

energy exchange is undetectable if the energy uncertainty $\Delta E$ of the electron beam is larger than $E_J$, as required in the above, in order to interact with the CPB at the right moment [41].

Here, in contrast to the discussion leading to equation (15), we treat the CPB quantum mechanically while the electron is viewed as a classical entity generating an external field. This is sufficient for the present purpose, although ultimately both the electron and the CPB should simultaneously be analyzed quantum mechanically. The Hamiltonian $H$ of the CPB near the charge degeneracy point is

$$H = 2E_C\rho (|0>_b<0|_b - |1>_b<1|_b) - (E_J/2)(|0>_b<1|_b + |1>_b<0|_b)$$
$$= (E_J/2)(|a>_b<a|_b - |s>_b<s|_b) + 2E_C\rho(|s>_b<a|_b + |a>_b<s|_b), \quad (16)$$

where $\rho = C_g V_g/e - 1$ is a dimensionless measure of the bias voltage that is zero at the charge degeneracy point [35]. The reflecting electron induces electrostatic polarization on the CPB and hence the effect of the electron can be regarded as an additional time-varying bias voltage $v_g(t)$ proportional to $\rho$. Equation (16) is analogous to the Hamiltonian of a two-state system appearing in, e. g. nuclear magnetic resonance textbooks. Let the time duration of electron-CPB interaction be $\tau_0$. As mentioned above, electron reflection flips the CPB state between $|s>_b$ and $|a>_b$. Hence, from the second line of equation (16) we see that, during the electron reflection, the magnitude of the product $\rho\tau_0$ is such that $2E_C\rho\tau_0/h \sim 1$ is satisfied. We intend to have a sufficiently short time duration $\tau_0$ so that $2E_C\rho \gg E_J/2$ is satisfied during the time duration. The CPB state precesses around the axis going through the states $|s>_b$ and $|a>_b$ on the Bloch sphere in the absence of the reflecting electron. In contrast, the CPB state precesses around the axis connecting the states $|0>_b$ and $|1>_b$ during the process of electron reflection because the term containing the factor $2E_C\rho$ in equation (16) dominates. Therefore, the transitions $|0>_b \leftrightarrow |1>_b$ is not induced during the short time interval of the latter precession (i. e. much shorter than the characteristic time $h/E_J$ for the former precession), as desired. Consequently, the reflection process is 'sudden' enough to ensure that the CPB states $|0>_b$ and $|1>_b$ remain the same, provided that the interaction time $\tau_0$ is sufficiently shorter than $h/E_J$.

## V. A MEASUREMENT PROTOCOL AND THE EXPECTED PERFORMANCE

### A. The protocol

First, we note that the relative phase between the CPB states $|s>_b$ and $|a>_b$ rotates with the frequency $E_J/h \sim 1$ GHz. In what follows, all the CPB-electron interactions and the CPB readout takes place when the phase factor $\exp(2\pi i E_J t/h)$ is 1 (for any fixed phase convention); and all the expressions will be written for such moments. This is the reason why the timing of electron emissions at the pulsed electron gun [28], together with the timing of microwave pulses for controlling the CPB [35], should have sub-ns precision. This precision needs to last for a time duration of perhaps $\sim 100$ ns, which depends on the measurement to be carried out, before single readout of the CPB state to obtain 1 bit of information.

Let us briefly go through the steps described in Sec. III. The first step is to initialize the CPB state to $|s>_b = (|0>_b + |1>_b)/2^{1/2}$. However, for later convenience we use a more general state

$$(|0>_b + g|1>_b)/2^{1/2}. \tag{17}$$

Assume for now that $|g| = 1$. After the first probe electron is reflected by the mirror, the electron and CPB is in an entangled state $(|0>_b|0> + g|1>_b|1>)/2^{1/2}$. The electron then interacts with the specimen. Suppose for now that inelastic scattering does not occur. Without loss of generality, we say that the electron state $|0>$ receives no phase shift, while the state $|1>$ acquires a phase factor $s = \exp(i\Delta\theta)$ during the interaction. Hence, after interacting with the specimen, the state becomes

$$(|0>_b|0> + gs|1>_b|1>)/2^{1/2}. \tag{18}$$

Next, we measure the state of the electron by the area detector ED (See Fig. 2). Let the electron state that corresponds to the $j$-th pixel of ED be $|j>_d$. Expand the states $|0>$ and $|1>$ in the eigenbasis $\{|j>_d| j = 1, 2, \ldots\}$, i. e. $|0> = \Sigma_j a_j |j>_d$ and $|1> = \Sigma_j b_j |j>_d$, where $\{a_j\}$ and $\{b_j\}$ are complex coefficients. Plugging them into equation (18), we see that after detecting the electron in the $j$-th pixel, the CPB is left in the state

$$(a_j|1>_b + b_j gs|0>_b)/(|a_j|^2 + |b_j|^2)^{1/2}. \tag{19}$$

At this point we introduce the *similar intensity map condition*: That is, the states $|0>$ and $|1>$ generate electron detection probability distributions on the ED such that $|a_j|$ and $|b_j|$ are nearly the same for all relevant $j$. If the similar intensity map condition is satisfied, we may write $b_j = a_j e^{i\beta_j}$ using some phase factor $\beta_j$ that depends solely on the electron optical design. Thus, the CPB state (19) can be written as, up to the overall phase factor,

$$(|0>_b + gse^{i\beta_j}|1>_b)/2^{1/2}. \tag{20}$$

Since we know $\beta_j$, the experimenter can correct the state (20) to obtain a state

$$(|0>_b + gs|1>_b)/2^{1/2}, \tag{21}$$

by applying a phase shift operation with the angle $-\beta_j$ on the CPB. (Alternatively, the experimenter may simply record all the experimental outcomes and perform the correction of the CPB state just before the final CPB state measurement step.) The overall effect is to get the state (20) from the initial state (17). As initially $g$ is 1, the CPB state becomes $(|0>_b + s^k|1>_b)/2^{1/2} = (|0>_b + e^{ik\Delta\theta}|1>_b)/2^{1/2}$ after $k$ electrons interact with the specimen.

The above process is followed by the final CPB readout step. We first apply to the CPB a phase shift operation with the angle $\pi/2$, which is followed by a single-qubit operation $|0>_b \rightarrow |s>_b$ and $|1>_b \rightarrow |a>_b$. This operation converts the phase $k\Delta\theta$ to the measurable amplitude, and is reminiscent of the use of a phase plate in phase contrast microscopy. The resultant state is, up to the overall phase factor

$$\{[\cos(k\Delta\theta/2) - \sin(k\Delta\theta/2)]|0>_b - i[\cos(k\Delta\theta/2) + \sin(k\Delta\theta/2)]|1>_b\}/2^{1/2}. \tag{22}$$

Finally, the CPB is measured with respect to the basis states $\{|0>_b, |1>_b\}$. The probabilities to find it in the states $|0>_b$ and $|1>_b$ are respectively $[1 - \sin(k\Delta\theta)]/2$ and $[1 + \sin(k\Delta\theta)]/2$. This result demonstrates in particular that the measurement is governed by the Heisenberg limit when $k \sim 1/\Delta\theta$. It is important to note that our protocol places a rather strict demand on the electron detector efficiency. If $k$ electrons are used before a single readout of the CPB state, the electron loss probability of the electron detector should be much less than $\sim 1/k$ [34].

A potential problem with the above protocol is robustness against electron inelastic scattering events. Such events, to be precise, entangle the quantum state of the electron with the state of the specimen; but it is more convenient to see these events as 'measurement of the electron state by the specimen'. Hence, the electron state is projected to a certain set of 'measurement basis states' after an inelastic scattering event. Since inelastic processes are known to produce a forward-scattered electron beam, it is generally accepted that inelastic scattering is rather delocalized and the associated length scale is on the order of several nanometers [36]. However, the matter could be more complex then this simple picture suggests, as there are reports of plasmon loss maps with sub-nanometer resolutions in the context of materials research [42].

We introduce the *delocalization hypothesis*, which we think is likely, though we do not claim to have a proof for its validity. That is, inelastic scattering is indeed delocalized over the region S0 and that the probabilities to find the electron in the state $|0>$ or $|1>$ right after an inelastic scattering event is still approximately 1/2. Hence we write the state, to which the electron state is projected, as

$$|\xi> = (|0> + e^{i\xi}|1>)/2^{1/2}, \qquad (23)$$

where $\xi$ is a phase parameter that may change from one inelastic scattering event to another. After this 'measurement' on the whole state (18), the CPB is left in the state

$$(|0>_b + gse^{-i\xi}|1>_b)/2^{1/2}. \qquad (24)$$

This expression has the same form as equation (20) and we may consider this inelastic scattering as having essentially the same effect on the CPB state as elastic scattering, except that we do not know the parameter $\xi$. However, the inelastic scattering angle $\Delta\theta_{inel} \sim 1$ mrad is generally smaller than the angle associated with elastic scattering. Hence, as a part of the delocalization hypothesis, we presume that the change of the angle of electron wave propagation is on the order of $\Delta\theta_{inel}$ after the 'measurement by the specimen'. Consequently, the experimenter may compute the parameter $\xi$ to the precision $\Delta\theta_{inel}$, by measuring the inelastically scattered electrons by the detector ED, in the same manner as computing the parameter $\beta_j$ by measuring the elastically scattered electrons. The above argument assumes that the chromatic aberration of the post-specimen lens system is made sufficiently small.

Finally, we mention issues that call for further investigations. One extreme possibility is that the electron state is projected onto the localized states $|0\rangle$ or $|1\rangle$ upon inelastic scattering. As can be seen from the state (18), such an event would fully destroy the CPB state of the form $(|1\rangle_b + g|0\rangle_b)/2^{1/2}$, making it impossible to continue the measurement. This possibility is actually unlikely because such localization would lead to wider inelastic scattering angles than what is observed experimentally. However, there is a possibility of *partially localized* inelastic scattering, after which the CPB state is of the form $a|1\rangle_b + b|0\rangle_b$, where the difference between the two amplitudes $|a|$, $|b|$ is no longer negligible. In fact, this problem also arises if the 'similar intensity map condition' mentioned above is partially violated. The difference between the amplitudes $|a|$, $|b|$ may then increase in the random-walk manner during the measurement, degrading the statistic in the final step of the CPB state measurement. Hence, this problem requires further study. There could also be a possibility to 'repair' the CPB quantum state to some extent by single-qubit operations on the CPB. Such a possibility also warrants further investigations.

## B. Estimating the performance by simulation

Monte-Carlo simulations were performed to evaluate the proposed method. The results are compared with simulated images of the best-case conventional electron microscopy, which is in-focus phase contrast microscopy with an ideally-working phase plate and a shot-noise limited electron detector. We use the ribosome molecule as an example specimen. Based on the atomic structure data of the ribosome obtained by X-ray crystallography [43], a phase shift map for 100 keV electrons is computed by using a formula described in Appendix A of Ref. [14] with zero-angle elastic scattering amplitude data provided in Ref. [49]. Figure 4 (a) shows the 30 nm x 30 nm phase shift map of the ribosome, which consists of 100 x 100 pixels. The image is smoothed with a Gaussian filter with a standard deviation 0.3 nm for better visibility.

In the proposed microscopy, we measure the difference $\Delta\theta$ of two phase shifts: i. e. one pertaining to the small region S1 and the other one pertaining to the surrounding region S0. We raster-scan both the incident beams |0> and |1> together, which correspond respectively to the regions S0 and S1. For each point during raster-scanning, the phase shift $\Delta\theta$ is recorded to form the resultant image. Hence, loosely speaking, we acquire a Laplacian-filtered phase shift map. We assume that the electron beam |0> is Gaussian with a standard deviation 1.5 nm, and likewise the beam |1> is Gaussian with a standard deviation 0.3 nm. To compute the phase shift map, which would be recorded if we were allowed to use an infinite electron dose, we calculated the difference of two phase maps: The first one is the ribosome phase shift map smoothed with a Gaussian filter with a standard deviation 0.3 nm (corresponding to the region S1). The second one is the same, except that the standard deviation of the Gaussian filter employed is 1.5 nm (corresponding to the region S0). The resultant map of the phase difference is shown in Fig. 4 (b). While high-resolution features are emphasized, low-resolution counterparts are diminished.

We performed Monte-Carlo simulations to see how the images produced by the proposed method would look like. We assume both the 'similar intensity map condition' and the 'delocalization hypothesis' to be valid. This is equivalent to assuming that the CPB state before each CPB state measurement is given by equation (22). Figures 4 (c) and (d) are the resultant images, whose grayscale corresponds linearly to the frequency of finding the CPB state in $|1>_b$ for each single measurement on the CPB. As in Fig. 4 (a), the images are smoothed with a Gaussian filter with a standard deviation 0.3 nm for better visibility. While the electron dose is 180 $e$/nm$^2$ for both the images, the number of electrons $k$ used before each single CPB measurement is 9 and 18, respectively. While we barely recognize the shape of the ribosome in Fig. 4 (c), some high-resolution features are clearly enhanced in Fig. 4 (d). Note, however, that the image is more 'grainy' for a larger $k$, reflecting the fact that a large $k$ is associated with a small number of the 'hypothetical particle', leading to large shot noise. These images represent 'raw data' of the proposed method; and how to process these data to obtain the best possible estimation of the structure of the specimen is a subject of future studies.

For comparison, we performed simulations to compute images by 'idealized' in-focus phase contrast microscopy at the same dose condition of 180 $e$/nm$^2$. For simplicity, we ignore inelastic scattering events. The probability to detect an electron in this setting is proportional to 1 + 2$\delta\theta$, where $\delta\theta$ represents the difference of phase shifts between the point of interest on the specimen and the average phase shift on the whole image [14]. The result is shown in Fig. 4 (e), in which the grayscale corresponds linearly to the electron count in each pixel, after applying a Gaussian filter with a standard deviation 0.3 nm. Comparing with Fig. 4 (c), we may say that the conventional in-focus phase contrast microscopy is superior to see low-resolution features. This is not surprising because the proposed method acquires Laplacian-filtered phase shift maps. To compare the two methods in the high-resolution region, we extracted high-resolution features from the electron count map in exactly the same way as we computed the image shown in Fig. 4 (b) from the X-ray phase shift map [45]. The result is shown in Fig. 4 (f), which shows less high-resolution features when compared with Fig. 4 (d).

## VI. CONCLUSION

We presented the followings. First, one of the major problems in biological cryoelectron microscopy, which is observation of weak phase objects with a limited number of electrons, is described. The use of quantum effects is identified as a possible route to go beyond the standard quantum limit. Second, the main idea of using quantum entanglement between the electron and the CPB for contrast enhancement is presented. Briefly, the idea involves linear accumulation of the weak phase shift caused by the specimen to the CPB. Third, experimental conditions to realize the proposed scheme have been examined. We basically argued that the realization of the microscope is possible with the presently available, or at least feasible, technologies. However, the new electron microscope would be highly unconventional, requiring a very narrow energy spread of the electron beam on the order of 0.1-1 meV within the portion of the microscope where the entangled electron quantum state is generated; although it does *not* entail necessity to stabilizing the high voltage source to such a degree. A very low temperature electron mirror working at ~ 10 mK, as well as sub-ns control of the electron pulses, are also involved in the microscope. Fourth, a measurement protocol to actually use the microscope is described; with an eye to inelastic processes that are unavoidable in biological electron microscopy. Numerical simulations have been performed to demonstrate the degree of contrast enhancement, under the assumption that delocalized inelastic scattering processes introduce only a small amount of amplitude and phase noise to the CPB, which we believe to be the case (See the description of the delocalization hypothesis in Sec. V). Finally, since the proposed method developed so far acquires differential phase shift maps (or Laplacian-filtered phase shift maps), it is suited only for obtaining high-resolution information, specifically at length scales smaller than the delocalization length. This seems to make the proposed method complementary to in-focus phase contrast microscopy, which is suited for acquiring structural information with small spatial frequencies.

## ACKNOWLEDGMENTS

The author thanks Professor Kazuyoshi Murata for discussions regarding in-focus cryoelectron microscopy, and Professor Li Xu for encouragement and support.

[45] We calculated the difference of two maps: The first one is the electron count map smoothed with a Gaussian filter with a standard deviation 0.3 nm shown in Fig. 4 (e). The second one is the same, except that the standard deviation of the Gaussian filter employed is 1.5 nm.

**Figure Captions**

FIG. 1. The concept of the proposed electron microscope. It contains an electron source (ES), electron mirror (EM), Cooper-pair box (CPB), specimen (SP), and area detector (AD). Two regions S0 and S1 within the specimen are also shown. Not all the lenses are shown in the figure.

FIG. 2. The proposed electron microscope. It consists of a pulsed electron source (PES), monochrometer (MC), electron beam separator (EBS), electron mirror (EM), Cooper pair box (CPB), condenser lens (CL), specimen holder (SH), objective lens (OL), projector lens system (PLS), and electron area detector (ED). Not all the lenses are shown in the figure.

FIG. 3. (a) A representative superconducting circuit containing a Cooper pair box (CPB), which is shown as the portion with thick lines. The CPB is connected to the ground electrode via a Josephson tunnel junction (JJ, the part surrounded by the dotted lines) with Josephson energy $E_J$ and the associated junction capacitance $C_J$. The CPB is also capacitively coupled via a coupling capacitor $C_g$ to a bias electrode held at the voltage $V_g$. (b) Configuration of the electrostatic electron mirror equipped with a CPB. Electrons are reflected in a manner that depends on the number of Cooper pairs in the CPB. (c) A diagram of electron rays. Electrons are first reflected by the mirror, then go through the lens system, followed by the specimen. (d) The electron mirror device may be seen as a quantum logic gate CNOT. This view can be used to estimate the effect of the electron on the CPB.

FIG. 4. Evaluation of the proposed method by computer simulation. (a) An electron phase shift map of the ribosome computed by using a known atomic structure obtained by X-ray

crystallography [43]. (b) The 'Laplacian-filtered' phase shift map mentioned in the text, which would be obtained by the proposed method if an unlimited electron dose were allowed. (c) An image obtained by simulating the proposed method with a finite electron dose 180 $e$/nm$^2$, with $k$ = 9 electrons associated with each single CPB measurement. (d) An image obtained under the same condition as image (c), except that the associated number of electrons is $k$ = 18. (e) A simulated image of conventional in-focus contrast electron microscopy with the same electron dose 180 $e$/nm$^2$. (f) High-resolution part of image (e) that is extracted for the purpose of comparison. The size of all images is (30 nm)$^2$.

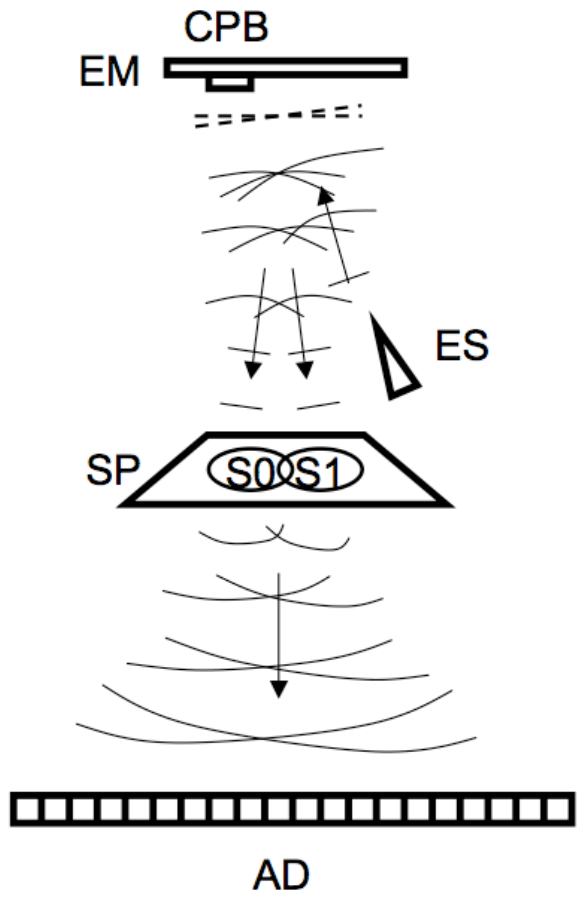

FIG. 1

Hiroshi Okamoto

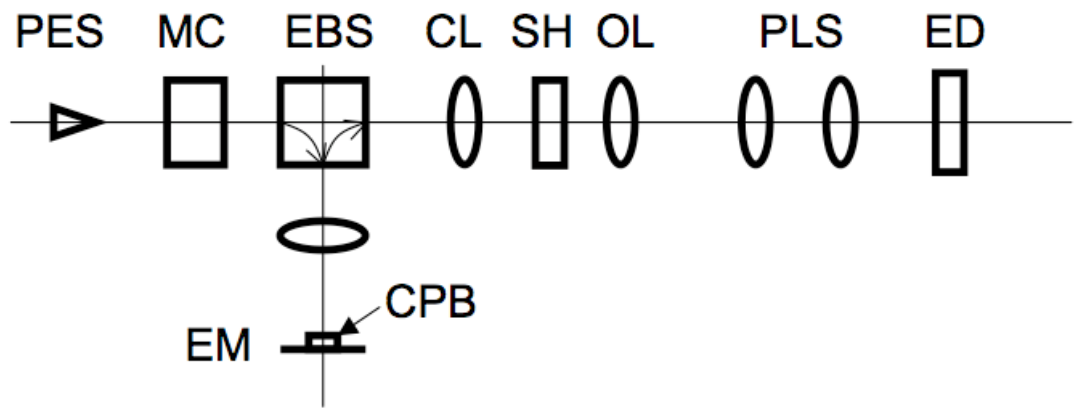

FIG. 2

Hiroshi Okamoto

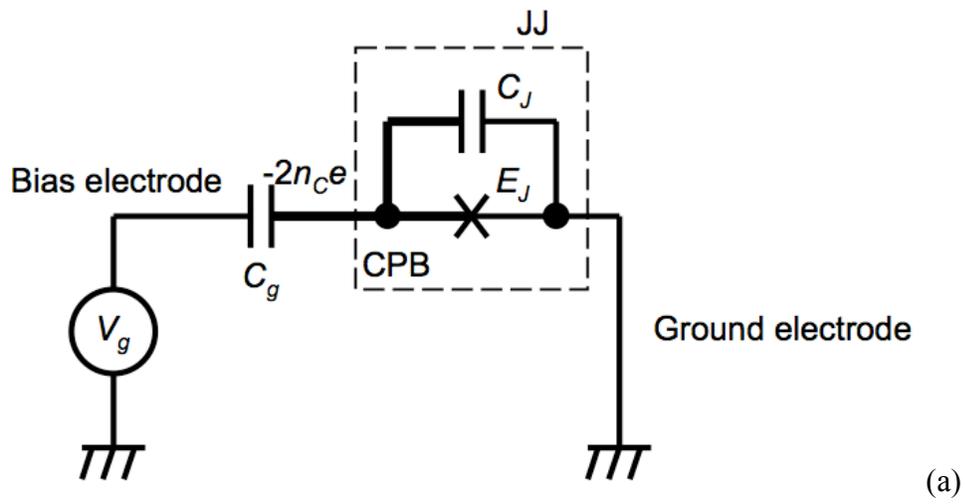

(a)

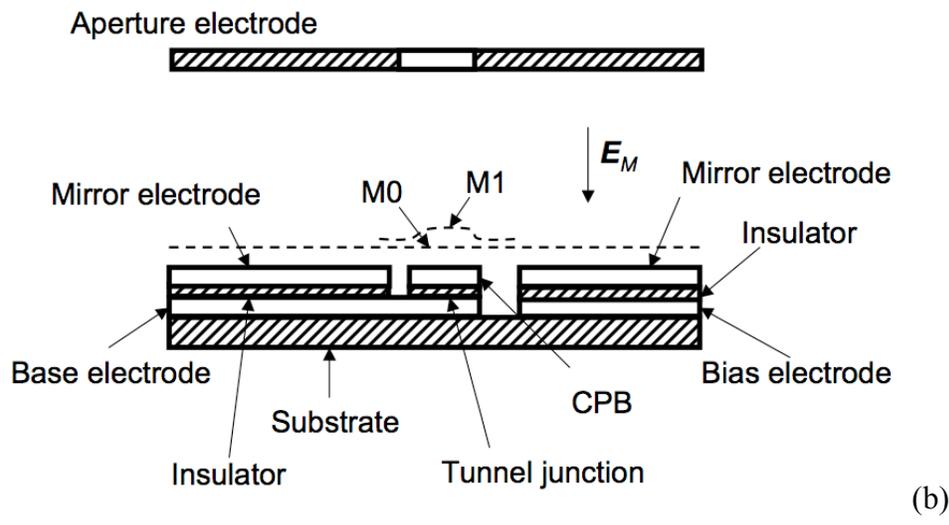

(b)

FIG. 3

Hiroshi Okamoto

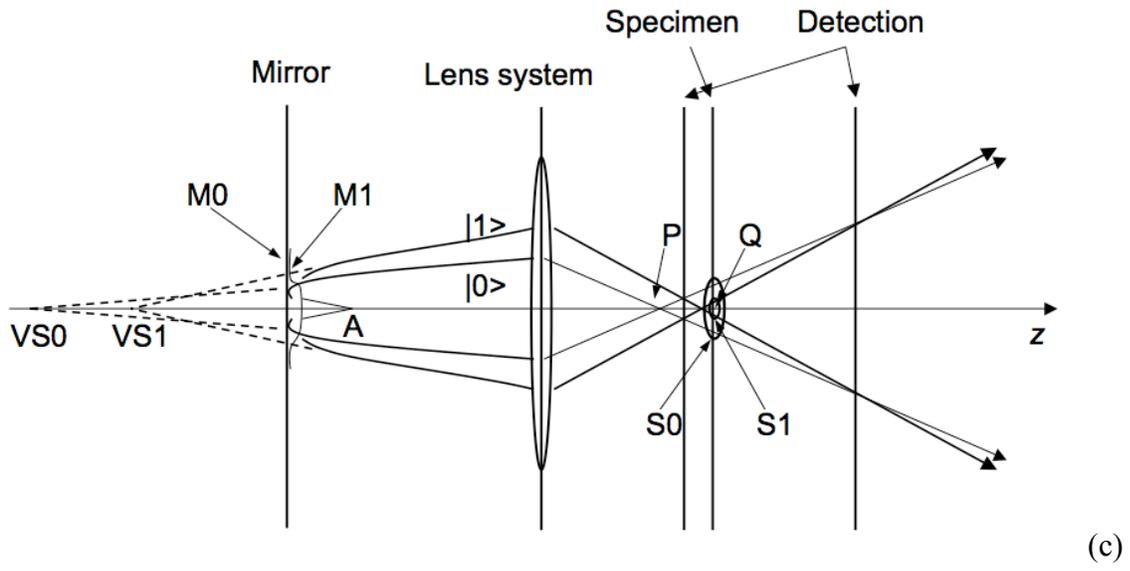

(c)

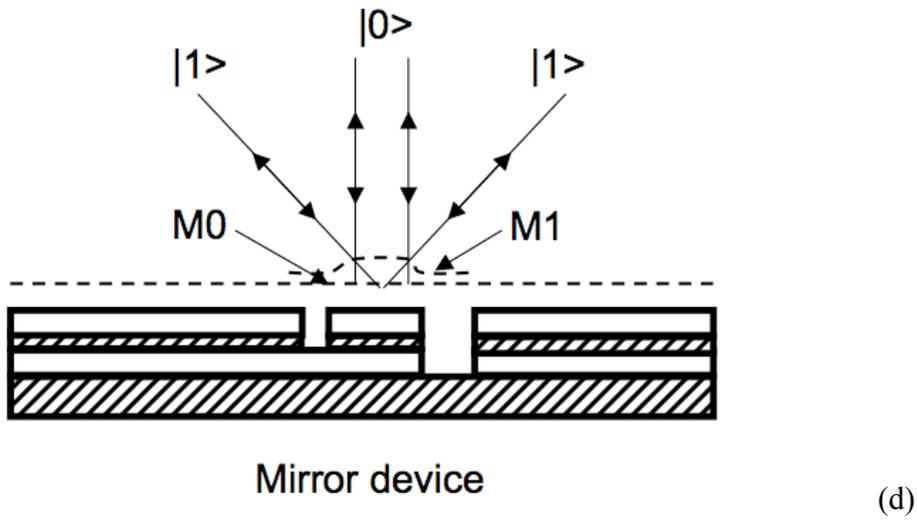

(d)

FIG. 3 (continued)

Hiroshi Okamoto

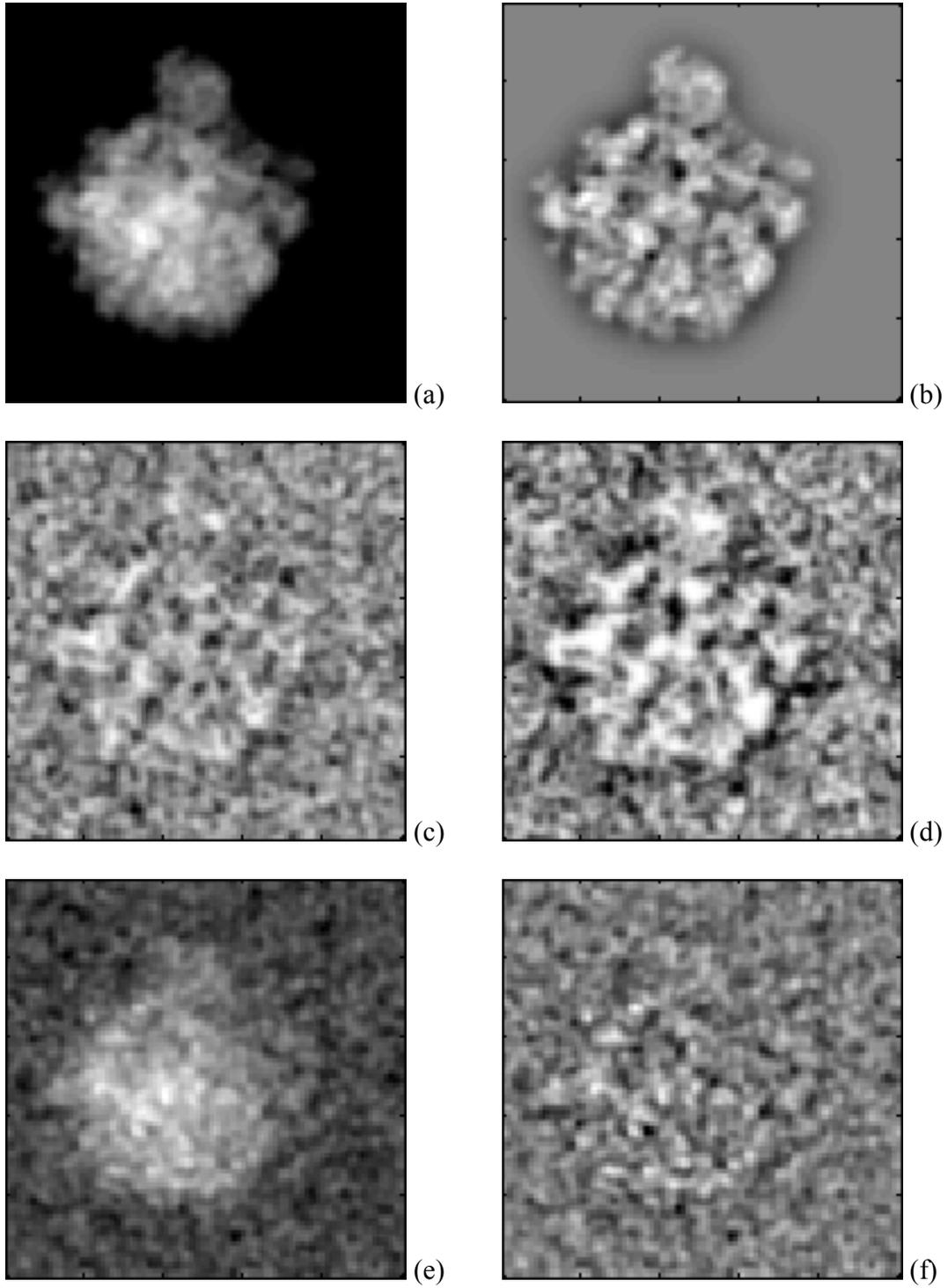

FIG. 4

Hiroshi Okamoto